\documentclass{rspublic}
\usepackage{amsmath,amssymb}
\usepackage{bm}

\newcommand{\eqdef}{\stackrel{\text{def}}{=}}
\newcommand{\n}{\nonumber \\}
\newcommand{\ignore}[1]{}

\newcommand{\Romannumeral}[1]{\uppercase\expandafter{\romannumeral#1}}

\newfont{\elevenmib}{cmmib10 scaled\magstep1}
\newcommand{\preprint}{
 \vspace*{-20mm}
    \begin{flushleft}
     \elevenmib Yukawa\, Institute\, Kyoto\\
   \end{flushleft}\vspace{-1.3cm}
   \begin{flushright}\normalsize \sf
     YITP-10-33\\
     April 2010
   \end{flushright}}

\allowdisplaybreaks[4]

\begin{document}


\newcommand{\Title}[1]{{\baselineskip=26pt
     \begin{center} \Large \bf #1 \\ \ \\ \end{center}}}
\newcommand{\Author}{\begin{center}
     \large \bf  Ryu Sasaki \end{center}}
\newcommand{\Address}{\begin{center}
             Yukawa Institute for Theoretical Physics,\\
       Kyoto University, Kyoto 606-8502, Japan
     \end{center}}
\newcommand{\Accepted}[1]{\begin{center}
     {\large \sf #1}\\ \vspace{1mm}{\small \sf Accepted for Publication}
     \end{center}}
 \preprint
\thispagestyle{empty}
\bigskip\bigskip

\Title{Exactly and Quasi-Exactly Solvable `Discrete'
Quantum Mechanics}
\Author

\Address
\vspace{1cm}

\begin{abstract}%
{}Brief introduction to the discrete quantum mechanics is given together with the
main results on various exactly solvable systems.
Namely, the intertwining relations, shape invariance, Heisenberg operator solutions,
annihilation/creation operators, dynamical symmetry algebras including the
$q$-oscillator algebra and the Askey-Wilson algebra.
A simple recipe to construct exactly and quasi-exactly
solvable Hamiltonians in one-dimensional `discrete' quantum mechanics
is presented.
It reproduces all the known ones whose eigenfunctions consist of
the Askey scheme of hypergeometric orthogonal polynomials of a
continuous or a discrete variable.
Several new exactly and quasi-exactly solvable ones are constructed. 
The sinusoidal coordinate plays an essential role.
\end{abstract}

\keywords{
{discrete quantum mechanics, difference Schr\"odinger equation, exact solvability, quasi-exact solvability, intertwining relations, shape invariance, Heisenberg operator solutions, closure relations, Askey-Wilson algebra}}

\section{Introduction}
\label{sec:intro}

This is an expanded version of my talk in Robin Bullough Memorial Symposium,
11 June 2009, Manchester.\footnote{Contribution to the Theme issue of the {\it Philosophical Transactions A} entitled Nonlinear phenomena, Optical and Quantum solitons, dedicated to the figure of Robin Bullough.}
I met Robin Bullough in the Niels Bohr Institute, Copenhagen in 1979, after finishing two papers on
unified geometric theory of soliton equations (Sasaki 1979a, b).
Robin was my mentor and navigator when I started cruising into an unknown sea
of integrable systems and soliton theory.
Thirty years ago, the subject was young and rough but it was very fertile and 
full of good promises. He gave me good suggestions and comments on my
third and fourth papers on geometric approach to soliton equations (Sasaki 1979c, 1980a). We produced joint papers, one on `Geometry of AKNS-ZS scheme' 
(Sasaki \& Bullough1980)
and the other on non-local conservation laws of sine-Gordon equation (Sasaki \& Bullough 1981).
Collaboration was a fruitful experience and I learned a lot from him and
got to know many friends and experts through him.
As a result, I am still working in the general area of integrable systems, and in particular, the general mechanism of integrability and solvability as a means to understand nonlinear interactions beyond perturbation, rather than some specific properties of
certain equations.

Here I present an overview of the problems I have been working in recent twelve years;
exactly and quasi-exactly solvable quantum particle dynamics. Due to the lack of space, 
I have to concentrate on the most basic but most interesting part of the
problems; the quantum mechanics (QM) of one degree of freedom.
The subjects to be covered in this article include: general setting of discrete quantum mechanics
with pure imaginary shifts and real shifts, shape invariance, exact solvability in the Heisenberg picture, dynamical symmetry algebras, unified theory of exactly and quasi-exactly solvable
discrete quantum mechanics. Let me emphasise that the basic structure of integrability and
methods of solutions are shared by the ordinary quantum mechanics (oQM) and the discrete 
quantum mechanics (dQM). 
For each specific subject, I will present basic formulas common to the ordinary and the discrete
QM and then provide explicit forms of various quantities which are different  in the oQM or
the two types of dQM.

One underlying motivation of this series of works is to inaugurate a new theory for the  Askey scheme of hypergeometric orthogonal polynomials (Andrews et al 1999; Nikiforov et al 1991; Ismail  2005; Koekoek \& Swarttouw 1996) within the framework of quantum mechanics, so that its abundant concepts and methods would be available for the research of orthogonal polynomials. 
So far this project has been extremely successful.
Every Hamiltonian system describing the hypergeometric orthogonal polynomial is shown to be
{\em shape invariant\/}  thus {\em exactly solvable in the Schr\"odinger picture\/} (Odake \& Sasaki 2005a, 2008b, c). As a byproduct a {\em universal Rodrigues type formula\/} 
\eqref{phin=A..Aphi0} for all the hypergeometric orthogonal polynomials is obtained.
It also turned out that all these  Hamiltonian systems describing the hypergeometric orthogonal polynomials are {\em exactly solvable in the Heisenberg picture\/} (Odake \& Sasaki 2006a, b).
Namely, the Heisenberg equation for the {\em sinusoidal coordinate\/} $\eta(x)$ is exactly solvable.
The positive/negative energy parts of the exact Heisenberg solution for $\eta(x)$ provide
the {\em annihilation/creation operators\/}, which together with the Hamiltonian constitute the {\em dynamical symmetry algebra\/} of the exactly solvable system. The {\em $q$-oscillator algebra\/} is obtained as a dynamical symmetry algebra of the system for the $q$-Hermite polynomials 
(Odake \& Sasaki 2008a). The relationship dictated by the annihilation/creation operators can be
interpreted as quantum mechanical disguise of the {\em three term recurrence relations\/}
of the orthogonal polynomials. Various {\em coherent states\/} as the eigenvectors of the
annihilation operators are explicitly obtained (Odake \& Sasaki 2006a, b, 2008c).
This project also opens a rich frontier for various {\em deformed polynomials\/} in terms of modification of Crum's theorem in the ordinary QM (Krein 1957; Adler 1994) and for the
discrete QM (Odake \& Sasaki 2010b). The problem of determining the deformed weight functions can be solved without recourse to the cumbersome moment problems.
One recent highlight of the project is the discovery of {\em infinitely many exceptional orthogonal
polynomials\/} related to the Laguerre, Jacobi, continuous Hahn, Wilson and Askey-Wilson polynomials (Odake \& Sasaki 2009c, d, e, 2010a, b; Ho,  Odake  \& Sasaki 2009).

\section{`Discrete' Quantum Mechanics}
\label{discQM}
\setcounter{equation}{0}

Throughout this paper we consider one-dimensional quantum mechanics. 
The dynamical variables are the coordinate $x$ and its conjugate momentum 
$p$, which is realised as a differential operator $p=-i\frac{d}{dx}\equiv -i\partial_x$.
The discrete quantum mechanics is a generalisation
of quantum mechanics in which the Schr\"{o}dinger equation is a
difference equation instead of differential in the ordinary QM
(Odake \& Sasaki 2008b, c). 
In other words, the Hamiltonian contains the momentum operator
 in exponentiated forms $e^{\pm\beta p}$ which
work as shift operators on the wavefunction
\begin{equation}
  e^{\pm\beta p}\psi(x)=\psi(x\mp i\beta).
\end{equation}
According to the two choices of the parameter $\beta$,
either {\em real\/} or {\em pure imaginary\/}, we have two types
of dQM;
with (\romannumeral1) pure imaginary shifts $\beta=\gamma\in\mathbb{R}_{\neq0}$ (pdQM), or (\romannumeral2)
real shifts $\beta=i$ (rdQM), respectively.
In the case of pdQM,
$\psi(x\mp i\gamma)$,  we require
the wavefunction to be an {\em analytic\/} function of $x$ with
its domain including the real axis or a part of it on which the
dynamical variable $x$ is defined. For the rdQM,
the difference equation gives constraints on wavefunctions only
on equally spaced lattice points. Then we choose, after proper
rescaling, the variable $x$ to be a non-negative  {\em integer\/}, with the
total number either finite ($N+1$) or infinite.

Here we will consider the Hamiltonians
having a finite or semi-infinite number of discrete energy levels only:
\begin{equation}
  0=\mathcal{E}(0)<\mathcal{E}(1)<\mathcal{E}(2)<\cdots.
  \label{positivesemi}
\end{equation}
The additive constant of the Hamiltonian is so chosen that
the groundstate energy vanishes. That is, the Hamiltonian is
{\em positive semi-definite}. It is a well known theorem in linear
algebra that any positive semi-definite hermitian matrix can be
factorised as a product of a certain matrix, say $\mathcal{A}$, and
its hermitian conjugate $\mathcal{A^\dagger}$.
As we will see shortly, the Hamiltonians of  one-dimensional quantum mechanics
have the same property, either the ordinary QM or the discrete QM both with the imaginary and real shifts.

\subsection{Hamiltonian} 
\label{sec:H}

The Hamiltonian of one dimensional  quantum mechanics has
a simple form 
\begin{equation}
\mathcal{H}\eqdef\mathcal{A}^\dagger \mathcal{A},
\end{equation}
in which the operators $\mathcal{A}$ and $\mathcal{A}^\dagger$ are:
\begin{align} 
\text{oQM}\quad  & \mathcal{A}\eqdef\frac{d}{dx}-\frac{dw(x)}{dx},\quad 
\mathcal{A}^\dagger=-\frac{d}{dx}-\frac{dw(x)}{dx},\quad w(x)\in\mathbb{R},
\label{oAdef}\\
\text{pdQM}\quad & \mathcal{A}\eqdef i\bigl(e^{\gamma p/2}\sqrt{V^*(x)}
  -e^{-\gamma p/2}\sqrt{V(x)}\bigr),\quad V(x),V^*(x)\in \mathbb{C},\n
  &\mathcal{A}^{\dagger}=-i\bigl(\sqrt{V(x)}\,e^{\gamma p/2}
  -\sqrt{V^*(x)}\,e^{-\gamma p/2}\bigr),\\
\text{rdQM}\quad &  \mathcal{A}\eqdef\sqrt{B(x)}-e^{\partial}\sqrt{D(x)},\qquad
B(x)\ge0,\quad D(x)\ge0,\n
  &\mathcal{A}^{\dagger}=\sqrt{B(x)}-\sqrt{D(x)}\,e^{-\partial},\quad  D(0)=0,\quad B(N)=0.
  \label{BDbound}
  \end{align}
  Here $V^*(x)$ is an analytic function of $x$ obtained from $V(x)$ by the $*$-operation,
which is defined as follows.
If $f(x)=\sum\limits_{n}a_nx^n$, $a_n\in\mathbb{C}$, then
$f^*(x)\eqdef\sum\limits_{n}a_n^*x^n$, in which $a_n^*$ is the complex
conjugation of $a_n$. Obviously $f^{**}(x)=f(x)$ and $f(x)^*=f^*(x^*)$.
If a function satisfies $f^*=f$, then it takes real values on the
real line.
As mentioned above, $e^{\pm\beta p}$ are shift operators
$e^{\pm\beta p}f(x)=f(x\mp i\beta)$, and the Schr\"odinger equation
\begin{equation}
  \mathcal{H}\phi_n(x)=\mathcal{E}(n)\phi_n(x),\qquad
  n=0,1,2,\ldots,
  \label{Sch_eq}
\end{equation}
is a second order differential (oQM) or difference equation (dQM).

Here are some explicit examples. 
{}From the ordinary QM. The {\em prepotential} $w(x)$ parametrises the groundstate wavefunction $\phi_0(x)$ which has no node and can be chosen real and positive $\phi_0(x)=e^{w(x)}$.
Then the above Hamiltonian takes the form $\mathcal{H}=p^2+U(x)$, $U(x)\eqdef (\partial_x w(x))^2+\partial_x^2w(x)$:
\begin{alignat}{2}
&\text{(\romannumeral1): Hermite}&\quad  {w}(x)&=-\tfrac12x^2, \hspace{30mm}-\infty<x<\infty,\n
  &{}&\quad U(x)&=x^2-1,\hspace{30mm} \eta(x)=x,
  \label{ex1}\\
&\text{(\romannumeral2): Laguerre} &\quad   {w}(x)&=-\tfrac12x^2+g\log x,\qquad\qquad g>1, \quad 0<x<\infty,
  \n
  &{}&\quad U(x)&=x^2+\frac{g(g-1)}{x^2}-1-2g,\qquad \eta(x)=x^2,\\
&\text{(\romannumeral3): Jacobi}&\quad  w(x)&=g\log\sin x+h\log\cos x,\quad g>1,\ h>1,\ 0<x<{\pi}/{2},
 \n
  &{}&\quad U(x)&=\frac{g(g-1)}{\sin^2x}+\frac{h(h-1)}{\cos^2x}-(g+h)^2, \quad \eta(x)=\cos2x.
\end{alignat}
{}From the discrete QM with pure imaginary shifts ($\beta=\gamma$, $0<q<1$):
\begin{alignat}{2}
&\text{(\romannumeral4): cont. Hahn}&{}&V(x)=(a_1+ix)(a_2+ix),\qquad \text{Re}(a_j)>0, \quad-\infty<x<\infty,\n
  &\qquad \gamma=1,&{}&\hspace{60mm} \eta(x)=x,
  \label{ex4}\\
&\text{(\romannumeral5): Wilson} &{}&V(x)=\frac{\prod_{j=1}^4(a_j+ix)}{2ix(2ix+1)},\qquad\qquad \text{Re}(a_j)>0,\quad 0<x<\infty,
  \n
  &\qquad \gamma=1,&{}&\{a_1^*,a_2^*,a_3^*,a_4^*\}=\{a_1,a_2,a_3,a_4\}\ \ (\text{as a set}),\ \ \eta(x)=x^2,\\
&\text{(\romannumeral6): Askey-Wilson}&\quad&V(x)=\frac{\prod_{j=1}^4(1-a_je^{ix})}
{(1-e^{2ix})(1-q\,e^{2ix})},\qquad |a_j|<1,\quad 0<x<\pi,
 \n
  &\qquad \gamma=\log q,&{}&\{a_1^*,a_2^*,a_3^*,a_4^*\}=\{a_1,a_2,a_3,a_4\}\ \ (\text{as a set}),\ \ \eta(x)=\cos x.
\end{alignat}
{}From the discrete QM with real shifts ($\beta=i$):
\begin{alignat}{2}
&\text{(\romannumeral7): Hahn}&{}&B(x)=(x+a)(N-x),\quad
  D(x)= x(b+N-x),\quad a>0, b>0,\n
 &\qquad &{}&\hspace{60mm} \eta(x)=x,\\  
&\text{(\romannumeral8): Racah}&{\quad}&B(x)=-\frac{(x+a)(x+b)(x+c)(x+d)}{(2x+d)(2x+1+d)},  \n
 &\qquad &{}&D(x)
  =-\frac{(x+d-a)(x+d-b)(x+d-c)x}{(2x-1+d)(2x+d)},\quad   \n
 &\qquad &{}&  c=-N,\quad a\ge b, \quad 0<b<1+d,\quad d>0,\quad a>N+d,\n
  &\qquad &{}&
 \tilde{d}\eqdef a+b+c-d-1,\qquad\qquad \eta(x)=x(x+d),\\
 &\text{(\romannumeral9): $q$-Racah}&{\quad}&B(x)
  =-\frac{(1-aq^x)(1-bq^x)(1-cq^x)(1-dq^x)}
  {(1-dq^{2x})(1-dq^{2x+1})}\,,\\[4pt]%
 &\qquad &{}&D(x)
=- \tilde{d}\,
  \frac{(1-a^{-1}dq^x)(1-b^{-1}dq^x)(1-c^{-1}dq^x)(1-q^x)}
  {(1-dq^{2x-1})(1-dq^{2x})}, \n
 &\qquad &{}& c=q^{-N},\ \ a\leq b,\ \ 0<d<1,\ \ 0<a<q^Nd,\  \ qd<b<1,\n
 &\qquad &{}& 
\tilde{d}<q^{-1},\ \ \tilde{d}\eqdef abcd^{-1}q^{-1},\qquad  \eta(x)=(q^{-x}-1)(1-dq^x).
 \label{ex9}
\end{alignat}
In the above list, the names like the Hermite, \ldots, $q$-Racah are the names of the eigenpolynomials of the corresponding Hamiltonian in the form
\begin{equation}
\phi_n(x)=\phi_0(x)P_n(\eta(x)),
\end{equation}
in which  $\eta(x)$ is called the {\em sinusoidal coordinate}. 
It is also given in the same list.
For the definitions and various properties of the hypergeometric orthogonal polynomials
and their $q$-versions in general, see (Koekoek \& Swarttouw 1996)  and 
(Odake \& Sasaki 2008b, c) in connection with dQM.

\bigskip
The groundstate wavefunction $\phi_0(x)$ is determined as a zero mode
of $\mathcal{A}$,
\begin{equation}
  \mathcal{A}\phi_0(x)=0.
  \label{Aphi0=0}
\end{equation}
The similarity transformed Hamiltonian $\widetilde{\mathcal{H}}$
in terms of the groundstate wavefunction $\phi_0(x)$ has a much simpler
form than the original Hamiltonian $\mathcal{H}$:
\begin{align}
  \widetilde{\mathcal{H}}&\eqdef
  \phi_0(x)^{-1}\circ \mathcal{H}\circ\phi_0(x)
  =\varepsilon\Bigl(V_+(x)(e^{\beta p}-1)+V_-(x)(e^{-\beta p}-1)\Bigr)
  \label{Htdef}\\
  &=\left\{
\begin{array}{lr}
\displaystyle{
-\frac{d^2}{dx^2}-2\frac{dw(x)}{dx}\frac{d}{dx}}  &   \text{oQM}   \\[4pt]
\displaystyle{\Bigl(V(x)(e^{\gamma p}-1)+V^*(x)(e^{-\gamma p}-1)\Bigr) } &   \text{pdQM}   \\[4pt]
\displaystyle{\Bigl(B(x)(1-e^{\partial})+D(x)(1-e^{-\partial})\Bigr)}  &   \text{pdQM}   
\end{array}
\right..
  \label{Ht}
\end{align}
It governs the differential (difference) equation of the polynomial eigenfunctions:
\begin{equation}
 \widetilde{\mathcal{H}}P_n(\eta(x))=\mathcal{E}(n)P_n(\eta(x)).
\end{equation}
Obviously, the square of the groundstate wavefunction $\phi_0(x)^2$
provides the orthogonality weight function for the polynomials:
\begin{align}
\text{oQM \& pdQM}\quad &\int_{x_1}^{x_2}\!\!\phi_0(x)^2P_n(\eta(x))P_n(\eta(x))dx\propto \delta_{n\,m},\\
\text{rdQM}\qquad\qquad
&\sum_{x=0}\phi_0(x)^2P_n(\eta(x))P_n(\eta(x))\propto \delta_{n\,m}.
\end{align}
Let me emphasise that the weight function, or $\phi_0(x)$ is determined as a
solution of a first order differential (difference) equation, without recourse
to a moment problem.
This situation becomes crucially important when various deformations of orthogonal polynomials are
considered. The explicit forms of the squared groundstate wavefunction  (weight function) 
$\phi_0(x)^2$ for the above examples in pure imaginary shifts dQM are:
\begin{align}
  &\phi_0(x)^2\n
  &=\left\{
  \begin{array}{ll}
 \displaystyle{ \Gamma(a_1+ix)\Gamma(a_2+ix)\Gamma(a_1^*-ix)\Gamma(a_2^*-ix)}
  &:\text{(\romannumeral4) cont. Hahn}\\[2pt]
  \displaystyle{(\Gamma(2ix)\Gamma(-2ix))^{-1}
  \prod_{j=1}^4\Gamma(a_j+ix)\Gamma(a_j-ix)}&:\text{(\romannumeral5) Wilson}\\[2pt]
  \displaystyle{(e^{2ix}\,;q)_{\infty}(e^{-2ix}\,;q)_{\infty}
  \prod_{j=1}^4(a_je^{ix}\,;q)_{\infty}^{-1}(a_je^{-ix}\,;q)_{\infty}^{-1}}
  &:\text{(\romannumeral6) Askey-Wilson}
  \end{array}\right.\!\!.
\end{align}
For the real shifts dQM, the zero-mode equation $\mathcal{A}\phi_0(x)=0$
  \eqref{Aphi0=0} is a two term recurrence relation, which can be solved elementarily by using
  the boundary condition \eqref{BDbound}:
\begin{align}
  \phi_0(x)^2&=\prod_{y=0}^{x-1}\frac{B(y)}{D(y+1)}\\
  &=\left\{
\begin{array}{ll}
\displaystyle{\frac{N!}{x!\,(N-x)!}\,\frac{(a)_x\,(b)_{N-x}}{(b)_N}}&   :\text{(\romannumeral7) Hahn}   
\\[4pt]
\displaystyle{\frac{(a,b,c,d)_x}{(1+d-a,1+d-b,1+d-c,1)_x}\,
  \frac{2x+d}{d}}&    :\text{(\romannumeral8) Racah}  \\[4pt]
\displaystyle{\frac{(a,b,c,d\,;q)_x}
  {(a^{-1}dq,b^{-1}dq,c^{-1}dq,q\,;q)_x\,\tilde{d}^x}\,
  \frac{1-dq^{2x}}{1-d}} &      :\text{(\romannumeral9) $q$-Racah}
\end{array}
\right..
  \label{dicsrete_phi0}
\end{align}
For the ordinary QM, the weight function is simply given by the prepotential $\phi_0(x)^2=e^{2w(x)}$.

\subsection{Intertwining Relations} 
\label{sec:Int}

Let us denote by $\mathcal{H}^{[0]}$ the  original factorised Hamiltonian and by 
$\mathcal{H}^{[1]}$ its partner Hamiltonian by changing the order of 
$\mathcal{A}^\dagger$ and $\mathcal{A}$:
\begin{equation}
\mathcal{H}^{[0]}\eqdef \mathcal{A}^\dagger\mathcal{A},\qquad
\mathcal{H}^{[1]}\eqdef \mathcal{A}\mathcal{A}^\dagger.
\end{equation}
One simple and most important consequence of the factorisation is the 
{\em intertwining relations\/}:
\begin{align}
\mathcal{A}\mathcal{H}^{[0]}=\mathcal{A}\mathcal{A}^\dagger\mathcal{A}
=\mathcal{H}^{[1]}\mathcal{A},\qquad
\mathcal{A}^\dagger\mathcal{H}^{[1]}=\mathcal{A}^\dagger\mathcal{A}\mathcal{A}^\dagger
=\mathcal{H}^{[0]}\mathcal{A}^\dagger,
\end{align}
which is equally valid in the ordinary and the discrete QM.
The pair of Hamiltonians $\mathcal{H}^{[0]}$ and $\mathcal{H}^{[1]}$ are essentially 
{\em iso-spectral\/} and their eigenfunctions $\{\phi_n^{[0]}(x)\}$ and $\{\phi_n^{[1]}(x)\}$
are related by the Darboux-Crum transformations (Darboux 1882; Crum 1955):
\begin{align}
\mathcal{H}^{[0]}\phi_n^{[0]}(x)&=\mathcal{E}(n)\phi_n^{[0]}(x),\hspace{40mm} n=0,1,\ldots,\\
\mathcal{H}^{[1]}\phi_n^{[1]}(x)&=\mathcal{E}(n)\phi_n^{[1]}(x),\hspace{40mm} n=1,2,\ldots,\\
\phi_n^{[1]}(x)&=\mathcal{A}\phi_n^{[0]}(x),\quad 
\phi_n^{[0]}(x)=\frac{\mathcal{A}^\dagger}{\mathcal{E}(n)}\phi_n^{[1]}(x),\qquad n=1,2,\ldots .
\end{align}
The partner Hamiltonian $\mathcal{H}^{[1]}$ has the lowest eigenvalue $\mathcal{E}(1)$.
If the groundstate energy $\mathcal{E}(1)$ is subtracted from the
partner Hamiltonian $\mathcal{H}^{[1]}$, it is again positive
semi-definite and can be factorised in terms of new operators $\mathcal{A}^{[1]}$ and 
$\mathcal{A}^{[1]\dagger}$:
\begin{equation}
\mathcal{H}^{[1]}\eqdef \mathcal{A}^{[1]\dagger}\mathcal{A}^{[1]}+\mathcal{E}(1).
\end{equation}
By changing the orders of $\mathcal{A}^{[1]\dagger}$  and  $\mathcal{A}^{[1]}$, a new Hamiltonian $\mathcal{H}^{[2]}$ is defined:
\begin{equation}
\mathcal{H}^{[2]}\eqdef\mathcal{A}^{[1]} \mathcal{A}^{[1]\dagger}+\mathcal{E}(1).
\end{equation}
These two Hamiltonians are intertwined by $\mathcal{A}^{[1]}$ and 
$\mathcal{A}^{[1]\dagger}$:
\begin{align}
\mathcal{A}^{[1]}(\mathcal{H}^{[1]}-\mathcal{E}(1))=\mathcal{A}^{[1]}\mathcal{A}^{[1]\dagger}\mathcal{A}^{[1]}
=(\mathcal{H}^{[2]}-\mathcal{E}(1))\mathcal{A}^{[1]},\\
\mathcal{A}^{[1]\dagger}(\mathcal{H}^{[2]}-\mathcal{E}(1))=\mathcal{A}^{[1]\dagger}\mathcal{A}^{[1]}\mathcal{A}^{[1]\dagger}
=(\mathcal{H}^{[1]}-\mathcal{E}(1))\mathcal{A}^{[1]\dagger}.
\end{align}
The iso-spectrality of the two Hamiltonians $\mathcal{H}^{[1]}$ and $\mathcal{H}^{[2]}$
and the relationship among their eigenfunctions follow as before:
\begin{align}
&\mathcal{H}^{[2]}\phi_n^{[2]}(x)=\mathcal{E}(n)\phi_n^{[2]}(x),\hspace{48mm} n=2,3,\ldots,\\
&\phi_n^{[2]}(x)=\mathcal{A}^{[1]}\phi_n^{[1]}(x),\quad 
\phi_n^{[1]}(x)=\frac{\mathcal{A}^{[1]\dagger}}{\mathcal{E}(n)-\mathcal{E}(1)}\phi_n^{[2]}(x),\qquad n=2,3,\ldots .
\end{align}
This process can go on indefinitely by successively deleting the lowest lying energy level:
\begin{align}
&\mathcal{H}^{[s]}\phi_n^{[s]}(x)=\mathcal{E}(n)\phi_n^{[s]}(x),\hspace{48mm} n=s,s+1,\ldots,\\
&\phi_n^{[s]}(x)=\mathcal{A}^{[s-1]}\phi_n^{[s-1]}(x),\quad 
\phi_n^{[s-1]}(x)=\frac{\mathcal{A}^{[s-1]\dagger}}{\mathcal{E}(n)-\mathcal{E}(s-1)}\phi_n^{[s]}(x),
\end{align}
for the ordinary QM (Crum 1955) and for the discrete QM as well (Gaillard \& Matveev 2009; Odake \& Sasaki 2009b).
The determinant expressions of the eigenfunctions $\phi_n^{[s]}(x)$ are also known for the
the ordinary QM (the Wronskian) and for the discrete QM as well (the Casoratian).

By {\em deleting\/} a finite number of energy levels from the original Hamiltonian systems
$\mathcal{H}^{[0]}$ and $\{\phi_n^{[0]}(x)\}$, instead of the successive lowest lying levels,
modification of Crum's theorem provides the essentially iso-spectral modified Hamiltonian $\bar{\mathcal H}$ and 
its eigenfunctions $\{\bar{\phi}_n(x)\}$. The set of deleted energy levels 
$\mathcal{D}=\{d_1,\ldots,d_\ell\}$ must satisfy the conditions
\begin{equation}
\prod_{j=1}^\ell(m-d_j)\ge0,\quad m\in\mathbb{Z}_+,
\end{equation}
in order to guarantee the hermiticity (self-adjointness) of the modified 
Hamiltonian $\bar{\mathcal H}$ and the non-singularity of the eigenfunctions 
\begin{equation}
\bar{\mathcal H}\bar{\phi}_n(x)=\mathcal{E}(n)\bar{\phi}_n(x),\qquad 
n\in\mathbb{Z}_+\backslash\mathcal{D}.
\end{equation}
Again the Wronskian (oQM) and Casoratian (pdQM) expressions of the 
eigenfunctions are known. 
For the ordinary QM, these results are known for some time (Krein 1957; Adler 1994).
For the dQM with pure imaginary shifts, the structure of the modified Crum's theory
was clarified only very recently (Garc\'ia-Guti\'errez, Odake \& Sasaki  2010).
Starting from an exactly solvable Hamiltonian, one can construct infinitely many  variants of
exactly solvable Hamiltonians and their eigenfunctions by Adler's and Garc\'ia et al's methods.
The resulting systems are, however, not shape invariant, even if the starting system is.
For the dQM with real shifts, some partial results are reported in (Yermolayeva \& Zhedanov 1999).

\subsection{Shape invariance} 
\label{sec:Sha}

Shape invariance (Gendenshtein 1983) is a sufficient condition for the exact solvability in
the Schr\"odinger picture. Combined with Crum's theorem (Crum 1955), or the factorisation method
(Infeld \& Hull 1951) or the so-called supersymmetric quantum mechanics (Cooper, Khare \& Sukhatme 1995), the totality of the discrete eigenvalues and the corresponding eigenfunctions
can be easily obtained. It was shown by Odake and Sasaki that the concept of shape invariance
works equally well in the discrete QM with the pure imaginary shifts (Odake \& Sasaki 2005a, 2008c)
as well as the real shifts (Odake \& Sasaki 2008b), providing quantum mechanical
explanation of the solvability of the Askey scheme of hypergeometric orthogonal polynomials in general.

In many cases the Hamiltonian contains some parameter(s),
$\bm{\lambda}=(\lambda_1,\lambda_2,\ldots)$.
Here we write parameter dependence explicitly, $\mathcal{H}(\bm{\lambda})$,
$\mathcal{A}(\bm{\lambda})$, $\mathcal{E}(n;\bm{\lambda})$,
$\phi_n(x;\bm{\lambda})$, etc, since it is the central issue.
The shape invariance condition is 
\begin{equation}
 \mathcal{A}(\bm{\lambda})\mathcal{A}(\bm{\lambda})^{\dagger}
  =\kappa\mathcal{A}(\bm{\lambda}+\bm{\delta})^{\dagger}
  \mathcal{A}(\bm{\lambda}+\bm{\delta})
  +\mathcal{E}(1;\bm{\lambda}),
  \label{shapeinv}
\end{equation}
where $\kappa$ is a real positive parameter and $\bm{\delta}$ is the
shift of the parameters. 
In other words $\mathcal{H}^{[0]}$ and $\mathcal{H}^{[1]}$ have the same shape, only the parameters are shifted by $\bm{\delta}$.
The energy spectrum and the excited state wavefunction are determined
by the data of the groundstate wavefunction $\phi_0(x;\bm{\lambda})$
and the energy of the first excited state $\mathcal{E}(1;\bm{\lambda})$
as follows:
\begin{align}
  &\mathcal{E}(n;\bm{\lambda})=\sum_{s=0}^{n-1}
  \kappa^s\mathcal{E}(1;\bm{\lambda}^{[s]}),\qquad\qquad\qquad\qquad \bm{\lambda}^{[s]}\eqdef
  \bm{\lambda}+{s}\bm{\delta},
  \label{shapeenery}\\
  &\phi_n(x;\bm{\lambda})\propto
  \mathcal{A}(\bm{\lambda}^{[0]})^{\dagger}
  \mathcal{A}(\bm{\lambda}^{[1]})^{\dagger}
  \mathcal{A}(\bm{\lambda}^{[2]})^{\dagger}
  \cdots
  \mathcal{A}(\bm{\lambda}^{[n-1]})^{\dagger}
  \phi_0(x;\bm{\lambda}^{[n]}).
  \label{phin=A..Aphi0}
\end{align}
The above formula for the eigenfunctions $\phi_n(x;\bm{\lambda})$ can be considered as the {\em universal Rodrigues type formula\/}
for the Askey scheme of hypergeometric polynomials and their $q$-analogues.
For the explicit forms of the Rodrigues type formula for each polynomial, one only has to
substitute the explicit forms of the operator $\mathcal{A}(\bm{\lambda})$ and the groundstate wavefunction 
$\phi_0(x;\bm{\lambda})$.
For the nine explicit examples given in \eqref{ex1}--\eqref{ex9}, it is straightforward to verify
the shape invariance conditions \eqref{shapeinv} and the energy \eqref{shapeenery} and
the eigenfunction \eqref{phin=A..Aphi0} formulas.
For the ordinary QM, the above shape invariance condition \eqref{shapeinv} becomes a relation
among the prepotentials ($\kappa=1$):
\begin{equation}
\left(\frac{dw(x;\bm{\lambda})}{dx}\right)^2-\frac{d^2w(x;\bm{\lambda})}{dx^2}=
\left(\frac{dw(x;\bm{\lambda}+\bm{\delta})}{dx}\right)^2+\frac{d^2w(x;\bm{\lambda}+\bm{\delta})}{dx^2}+\mathcal{E}(1;\bm{\lambda}),
\end{equation}
and the data for the three cases (\romannumeral1)--(\romannumeral3) are
\begin{alignat}{2}
&\text{(\romannumeral1): Hermite}&\quad  \bm{\lambda}&=\phi \ (\text{null}), 
\hspace{25mm} \mathcal{E}(n)=2n,\\
&\text{(\romannumeral2): Laguerre} &\quad   \bm{\lambda}&=g, \qquad\quad  \bm{\delta}=1, \qquad\qquad
  \mathcal{E}(n;\bm{\lambda})=4n,\\
&\text{(\romannumeral3): Jacobi}&\quad   \bm{\lambda}&=(g,h), \quad \bm{\delta}=(1, 1),
\qquad  \mathcal{E}(n;\bm{\lambda})=4n(n+g+h).
\end{alignat}
For the discrete QM with the pure imaginary shifts, the above shape invariance condition \eqref{shapeinv} is rewritten as:
\begin{align}
  &V(x-i\tfrac{\gamma}{2};\bm{\lambda})
  V^*(x-i\tfrac{\gamma}{2};\bm{\lambda})
  =\kappa^2\,V(x;\bm{\lambda}+\bm{\delta})V^*(x-i\gamma;\bm{\lambda}+\bm{\delta}),
  \label{contshape1}\\
  &V(x+i\tfrac{\gamma}{2};\bm{\lambda})
  +V^*(x-i\tfrac{\gamma}{2};\bm{\lambda})
  =\kappa\bigl(V(x;\bm{\lambda}+\bm{\delta})+V^*(x;\bm{\lambda}+\bm{\delta}))
  -\mathcal{E}(1;\bm{\lambda}).
  \label{contshape2}
\end{align}
The data for the three cases in pdQM  (\romannumeral4)--(\romannumeral6) are (Odake \& Sasaki 2005a, 2008c):
\begin{alignat}{3}
&\text{(\romannumeral4): cont. Hahn}&\quad  \bm{\lambda}&=(a_1,a_2), & &\bm{\delta}=(\tfrac12,\tfrac12), 
\qquad\quad \kappa=1,\n
&&\mathcal{E}(n;\bm{\lambda})&=4n(n+b_1-1),&  &b_1\eqdef a_1+a_2+a_1^*+a_2^*,\\
&\text{(\romannumeral5): Wilson} &\quad   \bm{\lambda}&=(a_1, a_2, a_3, a_4) &  &\bm{\delta}=(\tfrac12,\tfrac12,\tfrac12,\tfrac12), \quad
 \kappa=1,\n
&&\mathcal{E}(n;\bm{\lambda})&=4n(n+b_1-1),  &\quad & b_1\eqdef a_1+a_2+a_3+a_4,\\
&\text{(\romannumeral6): Askey-Wilson}&\quad   q^{\bm{\lambda}}&=(a_1, a_2, a_3, a_4), & &\bm{\delta}=(\tfrac12,\tfrac12,\tfrac12,\tfrac12), \quad  
\kappa=q^{-1},\n
&&\mathcal{E}(n;\bm{\lambda})&= (q^{-n}-1)(1-b_4q^{n-1}),& &b_4\eqdef a_1a_2a_3a_4.
\end{alignat}
%
%
For the discrete QM with the real shifts,
the shape invariance \eqref{shapeinv} is equivalent to the following
set of two equations:
\begin{align}
  &B(x+1;\bm{\lambda})D(x+1;\bm{\lambda})
  =\kappa^2\,B(x;\bm{\lambda}+\bm{\delta})D(x+1;\bm{\lambda}+\bm{\delta}),\\
  &B(x;\bm{\lambda})+D(x+1;\bm{\lambda})
  =\kappa\bigl(B(x;\bm{\lambda}+\bm{\delta})+D(x;\bm{\lambda}+\bm{\delta}))
  +\mathcal{E}(1;\bm{\lambda}).
\end{align}
The data for the three cases in rdQM  (\romannumeral7)--(\romannumeral9) are (Odake \& Sasaki 2008b):
\begin{alignat}{3}
&\text{(\romannumeral7): Hahn}&\quad  \bm{\lambda}&=(a, b, N), & &\bm{\delta}= (1, 1, -1)
\qquad\quad \kappa=1,\n
&&\mathcal{E}(n;\bm{\lambda})&=4n(n+a+b-1),&  &\\
&\text{(\romannumeral8): Racah} &\quad   \bm{\lambda}&=(a, b, d, N) &  &\bm{\delta}=(1, 1, 1, -1), \quad\ \
 \kappa=1,\n
&&\mathcal{E}(n;\bm{\lambda})&=4n(n+\tilde{d}),  &\quad & \\
&\text{(\romannumeral9): $q$-Racah}&\quad   q^{\bm{\lambda}}&=(a, b, d, q^{-N}), & &\bm{\delta}=(1, 1, 1, 1), \quad  
\kappa=q^{-1},\n
&&\mathcal{E}(n;\bm{\lambda})&= (q^{-n}-1)(1-\tilde{d}q^{n-1}).& &
\end{alignat}
It should be stressed that the size of the Hamiltonian ($N$ in the finite case) decreases by one, since the lowest eigenstate is removed.

\subsection{Solvability in the Heisenberg Picture }
\label{sec:Hei}

All the Hamiltonian systems describing the hypergeometric orthogonal polynomials 
are exactly solvable in the Heisenberg picture, too (Odake \& Sasaki 2006a, b). To be more precise, the Heisenberg operator of the sinusoidal operator $\eta(x)$
\begin{equation}
e^{it\mathcal{H}}\eta(x)e^{-it\mathcal{H}}
\end{equation}
can be evaluated in a closed form. The sufficient condition for that is the {\em closure relation\/}
\begin{equation}
  [\mathcal{H},[\mathcal{H},\eta(x)]\,]
  =\eta(x)\,R_0(\mathcal{H})+[\mathcal{H},\eta(x)]\,R_1(\mathcal{H})
  +R_{-1}(\mathcal{H}).
  \label{closurerel}
\end{equation}
Here $R_i(y)$ are polynomials in $y$. It is easy to see that the cubic commutator
$[\mathcal{H},[\mathcal{H},[\mathcal{H},\eta(x)]]]\equiv (\text{ad}\mathcal{H})^3\eta(x)$ is reduced to $\eta(x)$ and 
$[\mathcal{H},\eta(x)]$ with $\mathcal{H}$ depending coefficients:
\begin{align} 
(\text{ad}\mathcal{H})^3\eta(x)&= [\mathcal{H},\eta(x)]R_0(\mathcal{H})
+  [\mathcal{H},[\mathcal{H},\eta(x)]]\,R_1(\mathcal{H})\n
&=\eta(x)\,R_0(\mathcal{H})R_1(\mathcal{H})+[\mathcal{H},\eta(x)]\,(R_1(\mathcal{H})^2+R_0(\mathcal{H}))\n
&\qquad\qquad+R_{-1}(\mathcal{H})R_{1}(\mathcal{H}),
\end{align}
in which the definition $(\text{ad}\mathcal{H})X\eqdef [\mathcal{H}, X]$ is used.
It is trivial to see that all the higher commutators $(\text{ad}\mathcal{H})^n\eta(x)$ can also be reduced to $\eta(x)$ and 
$[\mathcal{H},\eta(x)]$ with $\mathcal{H}$ depending coefficients. Thus we arrive at
\begin{align}
     &e^{it\mathcal{H}}\eta(x)e^{-it\mathcal{H}}
=\sum_{n=0}^\infty\frac{(it)^n}{n!}({\rm ad}\,\mathcal{H})^n\eta(x),\n
  &=[\mathcal{H},\eta(x)]
  \frac{e^{i\alpha_+(\mathcal{H})t}-e^{i\alpha_-(\mathcal{H})t}}
  {\alpha_+(\mathcal{H})-\alpha_-(\mathcal{H})}
  -R_{-1}(\mathcal{H})/R_{0}(\mathcal{H})\n
  &\quad
  +\bigl(\eta(x)+R_{-1}(\mathcal{H})/R_0(\mathcal{H})\bigr)
  \frac{-\alpha_-(\mathcal{H})e^{i\alpha_+(\mathcal{H})t}
  +\alpha_+(\mathcal{H})e^{i\alpha_-(\mathcal{H})t}}
  {\alpha_+(\mathcal{H})-\alpha_-(\mathcal{H})},
  \label{quantsol}
\end{align}
in which the two ``frequencies" $\alpha_\pm(\mathcal{H})$ are
\begin{gather}
  \alpha_\pm(\mathcal{H})=\bigl(R_1(\mathcal{H})\pm
  \sqrt{R_1(\mathcal{H})^2+4R_0(\mathcal{H})}\,\bigr)/2, \\
  \alpha_+(\mathcal{H})+\alpha_-(\mathcal{H})=R_1(\mathcal{H}),
  \quad
  \alpha_+(\mathcal{H})\alpha_-(\mathcal{H})=-R_0(\mathcal{H}).
  \label{freqpm}
\end{gather}
The annihilation and
creation operators $a^{(\pm)}$ are extracted from this exact Heisenberg
operator solution:
\begin{align}
  &e^{it\mathcal{H}}\eta(x)e^{-it\mathcal{H}}
  =a^{(+)}e^{i\alpha_+(\mathcal{H})t}+a^{(-)}e^{i\alpha_-(\mathcal{H})t}
  -R_{-1}(\mathcal{H})R_0(\mathcal{H})^{-1},\\
    &a^{(\pm)}\eqdef\pm\Bigl([\mathcal{H},\eta(x)]-\bigl(\eta(x)
  +R_{-1}(\mathcal{H})R_0(\mathcal{H})^{-1}\bigr)\alpha_{\mp}(\mathcal{H})
  \Bigr)
  \bigl(\alpha_+(\mathcal{H})-\alpha_-(\mathcal{H})\bigr)^{-1}
  \label{a^{(pm)}}\n
  &\phantom{a^{(\pm)}}=
  \pm\bigl(\alpha_+(\mathcal{H})-\alpha_-(\mathcal{H})\bigr)^{-1}
  \Bigl([\mathcal{H},\eta(x)]+\alpha_{\pm}(\mathcal{H})\bigl(\eta(x)
  +R_{-1}(\mathcal{H})R_0(\mathcal{H})^{-1}\bigr)\Bigr).
\end{align}
The energy spectrum is determined by the over-determined recursion relations
$\mathcal{E}(n+1)=\mathcal{E}(n)+\alpha_+\bigl(\mathcal{E}(n)\bigr)$
and 
$\mathcal{E}(n-1)=\mathcal{E}(n)+\alpha_-\bigl(\mathcal{E}(n)\bigr)$
with $\mathcal{E}(0)=0$, and the excited state wavefunctions $\{\phi_n(x)\}$
are obtained by successive action of the creation operator $a^{(+)}$ on
the groundstate wavefunction $\phi_0(x)$.
This is the exact solvability in the Heisenberg picture.

The data for the three cases in oQM (\romannumeral1)--(\romannumeral3) are
\begin{alignat}{2}
&\text{(\romannumeral1): Hermite}&\quad  R_0(y)&=4, 
\hspace{10mm} R_1(y)=R_{-1}(y)=0,\\
&\text{(\romannumeral2): Laguerre} &\quad  R_0(y)&=16, \quad  
 R_1(y)=0, \quad R_{-1}(y)=-8(y+2g+1),\\
&\text{(\romannumeral3): Jacobi}&\quad   R_0(y)&=16(y+(g+h)^2), \quad 
R_1(y)=0, \n
&&& \hspace{30mm}
R_{-1}(y)=16(g-h)(g+h-1).
\end{alignat}
The data for the three cases in pdQM  (\romannumeral4)--(\romannumeral6) are (Odake \& Sasaki 2005a, 2008c):
\begin{align}
&\text{(\romannumeral4): cont. Hahn}\quad  R_0(y)=4y
  +4\text{Re}(a_1+a_2)\bigl(\text{Re}(a_1+a_2)-1\bigr), \quad R_1(y)=2, \n
&\phantom{\text{(\romannumeral4): cont. Hahn}}R_{-1}(y)=2\text{Im}(a_1+a_2)y
  +4\bigl(\text{Re}(a_1+a_2)-1\bigr)\text{Im}(a_1a_2),\\
&\text{(\romannumeral5): Wilson} \qquad R_0(y)=4y+b_1(b_1-2),\hspace{30mm} R_1(y)=2,\n
&\phantom{\text{(\romannumeral4): cont. Hahn}}R_{-1}(y)=-2y^2+(b_1-2b_2)y+(2-b_1)b_3,  \n
&\phantom{\text{(\romannumeral4): cont. Hahn}}b_2\eqdef\!\!\sum_{1\leq j<k\leq 4}a_ja_k,\quad
  b_3\eqdef\!\!\!\sum_{1\leq j<k<l\leq 4}a_ja_ka_l,\\
&\text{(\romannumeral6): Askey-Wilson}\quad R_1(y)=(q^{-\frac12}-q^{\frac12})^2y',
\qquad y'\eqdef y+1+q^{-1}b_4,\n
&\phantom{\text{(\romannumeral6): Askey-Wilson}}R_0(y)=(q^{-\frac12}-q^{\frac12})^2
  \bigl(y^{\prime\,2}-(1+q^{-1})^2b_4\bigr),\n
  &\qquad\quad R_{-1}(y)=-\tfrac12(q^{-\frac12}-q^{\frac12})^2
  \bigl((b_1+q^{-1}b_3)y'-(1+q^{-1})(b_3+q^{-1}b_1b_4)\bigr),\n
  &\phantom{\text{(\romannumeral6): Askey-Wilson}}b_1\eqdef\sum_{j=1}^4a_j,\quad
  b_3\eqdef\!\!\!\sum_{1\leq j<k<l\leq 4}a_ja_ka_l.
\end{align}
The data for the three cases in rdQM  (\romannumeral7)--(\romannumeral9) are (Odake \& Sasaki 2008b):
\begin{align}
&\text{(\romannumeral7): Hahn}\quad  R_0(y)=4y+(a+b-2)(a+b),\quad
R_1(y)=2,\n
&\phantom{\text{(\romannumeral7): Hahn}}R_{-1}(y)=-(2N-a+b)y-a(a+b-2)N,\\
&\text{(\romannumeral8): Racah} \quad   R_0(y)=4y+\tilde{d}^{\,2}-1, \quad R_1(y)=2,\n
&\phantom{\text{(\romannumeral8): Ra}}R_{-1}(y)=2y^2
  +\bigl(2(ab+bc+ca)-(1+d)(1+\tilde{d})\bigr)y
  +abc(\tilde{d}-1), \\
&\text{(\romannumeral9): $q$-Racah}\quad   R_0(y)=(q^{-\frac12}-q^{\frac12})^2
  \bigl(y^{\prime\,2}-(q^{-\frac12}+q^{\frac12})^2\tilde{d}\,\bigr),\quad y'\eqdef y+1+\tilde{d},\n
&\phantom{\text{(\romannumeral9): $q$-Racah}}\quad R_1(y)=(q^{-\frac12}-q^{\frac12})^2y',\n
&\quad R_{-1}(y)=(q^{-\frac12}-q^{\frac12})^2
  \Bigl((1+d)y^{\prime\,2}
  -\bigl(a+b+c+d+\tilde{d}+(ab+bc+ca)q^{-1}\bigr)y'\n
  &\qquad\qquad\qquad\qquad
  +\bigl((1-a)(1-b)(1-c)(1-\tilde{d}q^{-1})\n
  &\qquad\qquad\qquad\qquad
  +(a+b+c-1-d\tilde{d}+(ab+bc+ca)q^{-1})\bigr)(1+\tilde{d})\Bigr).
\end{align}

\subsection{Dual Closure Relation}
The {\em dual closure relation\/} has the same forms as the closure
relation \eqref{closurerel}  with the roles of the
Hamiltonian $\mathcal{H}$  and the
sinusoidal coordinate $\eta(x)$ exchanged:
\begin{align}
  [\eta,[\eta,\mathcal{H}]\,]&=\mathcal{H}\,R_0^{\text{dual}}(\eta)
  +[\eta,\mathcal{H}]\,R_1^{\text{dual}}(\eta)+R_{-1}^{\text{dual}}(\eta),
  \label{dualclosurerel}
\end{align}
in which
\begin{align}
  R_1^{\text{dual}}\bigl(\eta(x)\bigr)&=
  \bigl(\eta(x-i\beta)-\eta(x)\bigr)+\bigl(\eta(x+i\beta)-\eta(x)\bigr),
  \label{dualclcon1}\\
  R_0^{\text{dual}}\bigl(\eta(x)\bigr)&=
  -\bigl(\eta(x-i\beta)-\eta(x)\bigr)\bigl(\eta(x+i\beta)-\eta(x)\bigr),\\
  R_{-1}^{\text{dual}}\bigl(\eta(x)\bigr)&
  =\varepsilon\bigl(V_+(x)+V_-(x)\bigr)R_0^{\text{dual}}(\eta(x)).
  \label{dualclcon3}
\end{align}
The dual closure relation is the characteristic
feature shared by all the `Hamiltonians' $\widetilde{\mathcal{H}}$ which
map a polynomial in $\eta(x)$ into another.
Therefore its dynamical contents are not so constraining as the closure
relation, {\em except for\/} the real shifts (the discrete variable)
exactly solvable ($L=2$) case, where the closure relation and the dual
closure relations are on the same footing and they form a dynamical symmetry 
algebra which is sometimes called the Askey-Wilson algebra (Zhedanov 1992; 
Granovskii et al 1992; Terwilliger 2004; Koornwinder 2008;
Odake \& Sasaki 2008b).

\section{Unified Theory of Exact and Quasi-Exact Solvability}
\label{sec:Uni}
\setcounter{equation}{0}

In the previous section various examples of exactly solvable systems are explored and shown 
that the two sufficient conditions for exact solvability, the shape invariance and the closure
relation, are satisfied by them.
However, these two sufficient conditions
do not tell how to build exactly solvable models.
In this section I present a simple theory of constructing exactly
solvable Hamiltonians in discrete QM.
It covers all the known examples of exactly solvable discrete QM
with both pure imaginary and real shifts  and
it predicts several new ones to be explored.
The theory is general enough to generate
{\em quasi-exactly solvable\/} Hamiltonians in the same manner.
The quasi-exact solvability means, in contrast to the exact solvability,
that only a finite number of energy eigenvalues and the corresponding
eigenfunctions can be obtained exactly.
Many examples are known in the ordinary QM (Morozov et al 1990; Ushveridze 1994)
but only a few are known in dQM (Wiegmann \& Zabrodin 1994, 1995) .
This unified theory also incorporates the known examples of
quasi-exactly solvable Hamiltonians (Odake \& Sasaki 2007;
Sasaki 2007, 2008). 
A new type of quasi-exactly solvable Hamiltonians is constructed.
The present approach  reveals
the common structure underlying the exactly and quasi-exactly
solvable theories. This section is a brief introduction to a recent work
by Odake and Sasaki 2009a.

In the following I will take the similarity transformed Hamiltonian
$\widetilde{\mathcal{H}}$ \eqref{Htdef} instead of
$\mathcal{H}$ as the starting point. That is, I reverse the argument
and construct directly the `Hamiltonian' $\widetilde{\mathcal{H}}$
\eqref{Ht} based on the {\em sinusoidal coordinate\/} $\eta(x)$.
The general strategy is to construct the `Hamiltonian'
$\widetilde{\mathcal{H}}$ in such a way that
it maps a polynomial in $\eta(x)$ into another:
\begin{equation}
\widetilde{\mathcal{H}}\mathcal{V}_n \subseteq
\mathcal{V}_{n+L-2}\subset\mathcal{V}_{\infty}.
\end{equation}
Here $\mathcal{V}_n$ ($n\in\mathbb{Z}_{\geq 0}$) is defined by
\begin{equation}
  \mathcal{V}_n\eqdef
  \text{Span}\bigl[1,\eta(x),\ldots,\eta(x)^n\bigr], \qquad
  \mathcal{V}_{\infty}\eqdef\lim_{n\to\infty} \mathcal{V}_n.
  \label{Vndef}
\end{equation}

\subsection{potential functions}
\label{sec:pot_fn}

The  general form of the `Hamiltonian'
$\widetilde{\mathcal{H}}$ mapping a polynomial in $\eta(x)$  into another
is achieved by the following form of the potential functions
$V_{\pm}(x)$:
\begin{align}
  V_{\pm}(x)&=\frac{\widetilde{V}_{\pm}(x)}
  {\bigl(\eta(x\mp i\beta)-\eta(x)\bigr)
  \bigl(\eta(x\mp i\beta)-\eta(x\pm i\beta)\bigr)}\,,
  \label{V+-=V+-t/etaeta}\\[4pt]
  \widetilde{V}_{\pm}(x)&=\sum_{\genfrac{}{}{0pt}{}{k,l\geq 0}{k+l\leq L}}
  v_{k,l}\,\eta(x)^k\eta(x\mp i\beta)^l,
  \label{V+-t}
\end{align}
where $L$ is a natural number indicating the degree of $\eta(x)$ in 
$\widetilde{V}_{\pm}(x)$ and $v_{k,l}$ are real constants,
with the constraint $\sum\limits_{k+l=L}v_{k,l}^2\neq0$.
It is important that the same $v_{k,l}$ appears in both
$\widetilde{V}_{\pm}(x)$.
The `Hamiltonian' $\widetilde{\mathcal{H}}$ with the above $V_\pm(x)$
maps a degree $n$ polynomial in $\eta(x)$ to a degree $n+L-2$
polynomial.

The essential part of the formula \eqref{V+-=V+-t/etaeta} is the
denominators. They have the same form as the generic formula
 for the coefficients of the three term
recurrence relations of the orthogonal polynomials, (4.52) and (4.53)
in (Odake \& Sasaki 2008b).
The translation rules are the {\em duality\/} correspondence itself,
(3.14)--(3.18) in (Odake \& Sasaki 2008b):
\begin{gather}
  \mathcal{E}(n)\to\eta(x),\qquad -A_n\to V_+(x),
  \qquad -C_n\to V_-(x),\n
{}  \alpha_+\bigl(\mathcal{E}(n)\bigr)\to \eta(x-i\beta)-\eta(x),\quad
{}  \alpha_-\bigl(\mathcal{E}(n)\bigr)\to \eta(x+i\beta)-\eta(x).
\end{gather}

Some of the parameters $v_{k,l}$ in \eqref{V+-t} are redundant.
It is sufficient to keep $v_{k,l}$ with $l=0,1$.
The remaining $2L+1$ parameters $v_{k,l}$ ($k+l\leq L$, $l=0,1$) are
independent, with one of which corresponds to the overall normalisation
of the Hamiltonian.

The $L=2$ case is exactly solvable.
Since the Hamiltonian of the polynomial space $\widetilde{\mathcal{H}}$
is an upper triangular matrix, its eigenvalues and
eigenvectors are easily obtained explicitly.
For the solutions of a full quantum mechanical problem, however,
one needs the square-integrable groundstate wavefunction $\phi_0(x)$
\eqref{Aphi0=0}, which is essential for the existence of the Hamiltonian
$\mathcal{H}$ and the verification of its hermiticity. These conditions
would usually restrict the ranges of the parameters $v_{0,0},\ldots,v_{2,0}$.
It is easy to verify that the explicit examples of the potential functions $V(x)$,
$V^*(x)$ and $B(x)$, $D(x)$ in \eqref{ex4}--\eqref{ex9} are simply reproduced by proper choices of the 
parameters $\{v_{k,l}\}$. It should be stressed that the above form of the potential function 
\eqref{V+-=V+-t/etaeta}--\eqref{V+-t} provides a {\em unified proof\/} of 
the shape invariance relation \eqref{shapeinv}, the closure relation \eqref{closurerel}
and the dual closure relation \eqref{dualclosurerel} in the $\widetilde{\mathcal H}$ scheme.

The higher $L\ge3$ cases are obviously non-solvable. Among them, the tame non-solvability of $L=3$ and $4$ can be made quasi-exactly solvable (QES) by adding suitable compensation terms. This is a simple generalisation of the method of Sasaki \& Takasaki 2001 for multi-particle QES in
the ordinary QM.
For a given positive integer $M$, let us try to find a QES `Hamiltonian'
$\widetilde{\mathcal{H}}$, or more precisely its modification
$\widetilde{\mathcal{H}}'$, having an invariant polynomial subspace
$\mathcal{V}_M$:
\begin{equation}
  \widetilde{\mathcal{H}}'\mathcal{V}_M\subseteq\mathcal{V}_M.
\end{equation}
For $L=3$, $\widetilde{\mathcal{H}}'$ is defined by adding
one single compensation term  of degree one
\begin{equation}
  \widetilde{\mathcal{H}}'\eqdef\widetilde{\mathcal{H}}-e_0(M)\eta(x),
    \label{L3ham}
\end{equation}
and we have achieved the quasi-exact solvability
$\widetilde{\mathcal{H}}'\mathcal{V}_M\subseteq\mathcal{V}_M$.
Known discrete QES examples belong to this class (Sasaki 2007, 2008).

For $L=4$ case, $\widetilde{\mathcal{H}}'$ is defined by adding a
linear and a quadratic in $\eta(x)$ compensation terms to the
Hamiltonian $\widetilde{\mathcal{H}}$:
\begin{equation}
  \widetilde{\mathcal{H}}'\eqdef
  \widetilde{\mathcal{H}}-e_0(M)\eta(x)^2-e_1(M)\eta(x),
    \label{L4ham}
\end{equation}
and $\widetilde{\mathcal{H}}'\eta(x)^M\in\mathcal{V}_M$.
This type of QES theory is new.
\section*{Acknowledgements}
R.\,S. is supported in part by Grant-in-Aid for Scientific Research from
the Ministry of Education, Culture, Sports, Science and Technology,
No.22540186.


\end{document}